# NACS: non-overlapping AP's caching scheme to reduce handoff in 802.11 wireless LAN

Usman Tariq, Yasir Malik, Man-Pyo Hong

Graduate School of Information and Communication, Ajou University
San 5, Wonchon-Dong, Youngtong-Gu, Suwon, Kyonggi-Do, 443-749, Korea
{usman, yasir, mphong,}@ajou.ac.kr

*Abstract* - With the escalation of the IEEE 802.11 based wireless networks, Voice over IP and analogous applications are also used over wireless networks. Recently, the wireless LAN systems are spaciously deployed for public internet services. In public wireless LAN systems, reliable user authentication and mobility support are indispensable issues. When a mobile device budges out the range of one access point (AP) and endeavor to connect to new AP, it performs handoff. Contemporarily, PNC and SNC were proposed to propagate the MN context to the entire neighboring AP's on the wireless network with the help of neighbor graph. In this paper, we proposed a non-overlapping AP's caching scheme (NACS), which propagates the mobile node context to those AP's which do not overlap with the current AP. To capture the topology of non-overlapping AP's in the wireless network, non-overlapping graph (NOG) is generated at each AP. Simulation results shows that NACS reduces the signaling cost of propagating the MN context to the neighbor AP's in the wireless network.

## 1 Introduction

Recently, the wireless LAN (WLAN) systems are spaciously deployed for the public internet services. Most of current WLAN networks for the internet access are particularly based on IEEE 802.11 standards [1] providing connectivity up to 11 Mbps to 54 Mbps. In the beginning, IEEE 802.11 was originally designed for the indoor networks where hosts were stationary and mobility was not an issue [2]. However, with the rapid growth of wireless networks and portable devices, mobility support in IEEE 802 becomes one of the most important issues to be solved.

Since there is lack of mobility support in IEEE 802.11 standards, it causes a major interruption while performing handoff. Fig.1. shows a typical topology of handoff in a WLAN network. A handoff occurs when a mobile node moves from the radio range of current access point to the next potential access point.

The handoff procedure consists of scanning, authentication and re-association [3]. Current Wi-Fi-based networks are not supported by the layer 2 handoff latencies and contribute approximately 90% of the total latency which exceeds 100ms [2][4]. We know that for good voice over IP service, the handoff latency must not exceed 53 ms [5].

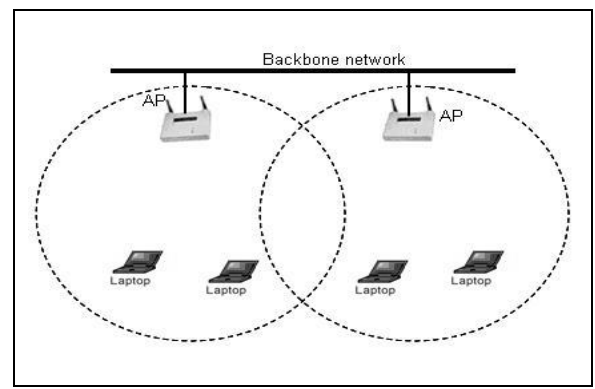

**Fig.1.** Typical Topology for Handoff in WLAN

To reduce the handoff latency in IEEE 802.11 wireless network, many researchers have proposed different techniques to reduce the layer 2 and layer 3 handoff latencies. In [3][5][6] authors study and suggest various techniques to reduce the layer 2 latency and accomplished to reduce the latencies up to 20ms to 60ms depending on the client.

Handoff involves transfer of mobile node context information from old access point to new access point [7]. The context transfer is done via inter-access point communication. Consequently, there is some latency involve in this process which increase the overall handoff latency.

Pervious research in this perspective was focused on transfer of context information in a reactive fashion, which implies that the transfer of context will be initiated when a mobile node will associate with the new access point which causes increase in the overall handoff latency rather then reducing it. One method to reduce handoff latency is to transfer the mobile node context ahead of mobile device in proactive fashion.

Proactive neighbor caching [8] scheme was proposed to reduce the context transfer latency. PNC scheme transfer the context ahead of mobile node in a proactive manner. To determine the potential next AP, it uses the neighbor graph which dynamically captures the topological information of the wireless network. When neighbor AP is determined it transfer the mobile node context to it in advance. Currently PNC scheme is included in the Inter access point protocol (IAPP) specification.

Sangheon Pack proposed selective neighbor caching (SNC) [9] which is quite similar to PNC. An AP in SNC

This research is supported by the ubiquitous Autonomic Computing and Network Project, the Ministry of Information and Communication (MIC) 21st Century Frontier R&D Program in Korea.

scheme, proactively propagate the MN context to the neighbor AP's whose handoff probabilities are equal to or higher then a predefined threshold value. The optimal performance of SNC scheme depends on the value of threshold, which has to be carefully determined.

In both schemes, the signaling cost involves in transferring the mobile node context information was high. There is directly proportional relationship between the number of mobile nodes in the network and the signaling cost for transferring the context information i.e. if the number of nodes increases in the network, signaling cost for transferring the context also increases.

In this paper, we proposed a new scheme non-overlapping AP caching scheme (NACS) to reduce the context transfer latency. In NACS the context is transmitted to the APs which do not overlap to each other. To capture the topology of non overlapping neighbor AP's in a wireless network, a non overlapping graph is generated at each AP. A non overlapping graph can be generated much faster than the neighbor graph.

The rest of paper is organized as follows. In section 2, we describe the handoff process in IEEE 802.11 networks. Section 3 describes the related work for the context caching. Section 4, describes the non-overlapping AP's caching scheme. Section 5 shows the simulation results in terms of cache hit probability and signaling cost; finally we conclude our work in section 6.

## 2 The handoff Process

Two types of operations are allowed in 802.11 media access specification: adhoc and infrastructure mode [4]. In adhoc mode, two or more mobile nodes make peer to peer relationship after recognizing each other. While in infrastructure mode mobile node can communicate with other mobile nodes only when there is a base station exists. Thus it is necessary for a mobile node to connect with any base station to communicate with the other mobile nodes.

A handoff occurs when a mobile device moves beyond the radio range of one access point and enters to another access point. Handoff process involves various MAC and network layer functions.

There are two distinct logical types of handoff: first is discovery and second one is re-authentication. Following are the steps to complete the hand off process:
1) Probing: The best access point discovery is important issue in this process.
2) Authentication: Two type authentications are necessary to authenticate the mobile nodes of network: Open authentication and WEP (Wired Equivalent Privacy).
3) Re-association: Link establishment with new access point.

Usually in WLAN handoff latencies occurs at layer 2 and layer 3. In this paper, our focus was only on reducing layer 2 handoff latency. During the handoff phase management frames are transferred between mobile nodes, access point and between *Old AP* and *New AP*. Fig 2 shows the sequence of messages which are exchanged between mobile node and access point. It is assumed that the mobile node has been terminated by some access point to which it was connected earlier. Fig.2. shows the sequence of messages involved after terminating from the current AP to which mobile node was connected.

In Fig.2. mobile node is referring to a device capable of performing its participation as an 802.11 mobile node. *Old AP* is an access point with which the mobile device is associated before handoff procedure and *New AP* is the access point through which the mobile node will be connected after performing handoff.

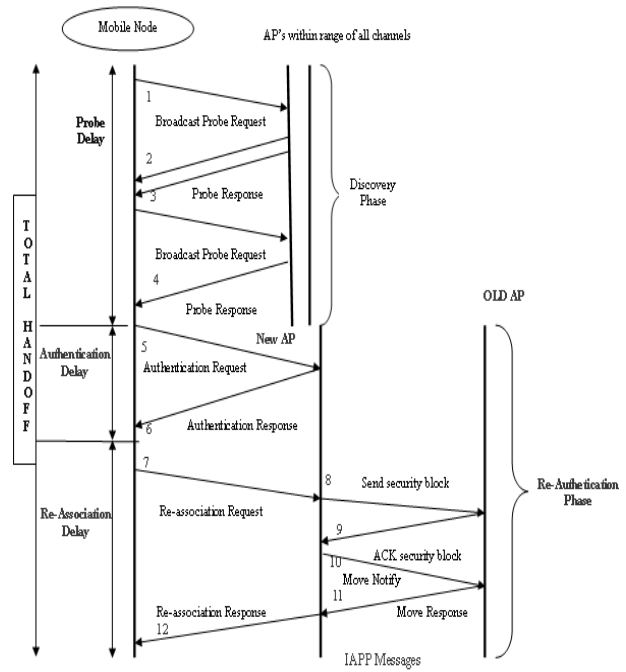

**Fig.2.** The Handoff procedure in 802.11 and 802.11f

In fig.2 the messages '1' through '4' shows the mobile node scans for the access point. Scanning can be performed by sending the probe request which is said to be active scanning or by listening to the becon messages which is said to be passive scanning. Messages 1 to 4 shows the active scanning, mobile node sends probe request in message 1 and 3 and receives probe response from the APs in message 3 and 4. After scanning mobile node select the new access point on the basis of signal strength. To be authenticated with the selected AP mobile node sends the authentication request in message '5' after authenticating access point send the authentication response in messages 6. After authentication, mobile node send re-association request to new AP. At message 7, the mobile node sends the re-association request to the new AP and receives the response at message 12 to complete the handoff procedure. The delay during the request and response of re-association messages is called the re-association latency. During this delay mobile node context information is exchanged between new and old AP via inter access point protocol (IAPP). Message 8 to 11 shows the message exchanges between new and old access point.

## 2.1 IAPP [Inter Access Point Protocol]

IAPP (Inter Access-Point Protocol) [10] or IEEE Standard 802.11f is a part of IEEE 802.11, designed for the enforcement of unique association throughout a ESS (Extended Service Set) and for secure exchange of mobile node's security context between current access point(AP) and new AP during handoff period. Based on security level, communication session keys between AP's are distributed by a RADIUS server. RADIUS server also provides a mapping service between AP's MAC address and IP address. The new AP sends security block for old AP which it received from the RADIUS server, as a send security block packet. This is first message in IAPP TCP exchange between the AP's. The old access point returns the acknowledgement security block packet. Both AP's now have to encrypt all the packets for the exchange between the AP's.

In inter access point protocol, the context is sent from the IAPP entities to each neighbor AP in the CACHE notify request expires. The RADIUS server investigates that the AP is an authenticated member of ESS, when it gets receipt of registration access request from the AP. If the radius server determines that AP is not the authenticated member of ESS, the RADIUS server will reply to AP's registration access request packets with RDIUS registration access reject. The IAPP entity should delete the neighboring AP's that did not respond before the expiration of the time out of the neighboring graph.

## 3 Related Work

### 3.1. PNC

Before implementing the neighbor graph, the link layer context caching was originally reactive. Later PNC scheme was proposed, it works proactively and propagates the context information one hop ahead of mobile node. Fig.3. shows the basic operation of PNC scheme, initially neighbor graph is constructed on all AP's which capture the topological information of the wireless network. The AP located in center is the current AP to which mobile is connected initially. When a mobile node is connected to the AP, its context information is propagated to all neighbor APs based on the neighbor graph.

In PNC all neighboring AP's receives the context information of mobile node. When ever a mobile node attached to an AP, AP propagates its context information prior to its movement to all neighboring APs. Here mobile node context means information regarding mobile node session, QoS and security [11].

Following notations are used to describe the proactive neighbor caching scheme algorithm.

1) *Context (c):* represents the context information of a client 'c'
2) *Cache ($ap_m$):* represent the maintained data structure of cache at $ap_m$.
3) *Propagate_Context ($ap_l$, c, $ap_m$):* represents the context propagation of client c from $ap_l$ to $ap_m$
4) *Obtain Context($ap_l$, c, $ap_m$)* used when $ap_m$ obtain context(c) from $ap_l$ using IAPP Move_notify massage
5) *Remove_Context ($ap_l$, c, $ap_m$)* here $ap_l$ sends a chche_invalidate message to $ap_m$ to remove the context(c) from cache(m)

---
**Algorithm 1** PNC Algorithm ($ap_m$, c, $ap_l$)
---
1: $ap_m$: the current-AP, $ap_l$: the old AP, c: the client;
2: **if** client c associates to $ap_m$ **then**
3:   **for all** $ap_l \in$ Neighbor($ap_m$) **do**
4:     Propagate Context($ap_m$, c, $ap_l$)
5:   **end for**
6: **end if**
7: **if** client c re-associates to $ap_m$ from $ap_n$ **then**
8:   **if** Context(c) not in Cache($ap_m$) **then**
9:     Obtain Context($ap_n$, c, $ap_m$)
10:   **end if**
11:   **for all** $ap_l \in$ Neighbor ($ap_m$) **do**
12:     Propagate Context($ap_m$, c, $ap_l$)
13:   **end for**
14: **end if**
15: **if** client c re-associates to $ap_n$ from $ap_m$ **then**
16:   **for all** $ap_l \in$ Neighbor ($ap_m$) **do**
17:     Remove Context($ap_m$, c, $ap_l$)
18:   **end for**
19: **end if**
20: **if** $ap_m$ received Context(c) from $ap_l$ **then**
21:   Insert Cache($ap_m$ ,Context(c))
22: **end if**
---

PNC scheme algorithm is presented in Algorithm 1. In PNC algorithm, when a MN associates to an AP, it will propagate its context to all neighbors AP's (lines 2-6). If the context information is not found in cache of new neighboring AP, it will request to old AP for the context information of the respective mobile node. After receiving the context information the $AP_l$ propagate the context information to all neighboring AP's (lines 7-14). After context is transferred, old AP and neighboring AP remove mobile node context form its cache. (15-19). When context information is received form the AP it is inserted in AP cache (line 20-22).

### 3.2. SNC

Second most adopted scheme to reduce handoff is SNC (Selective Neighbor Cache Scheme). As shown in fig 3 neighboring APs of the current AP which is located in the center have different handoff probabilities which are calculated with the similar method in [12].

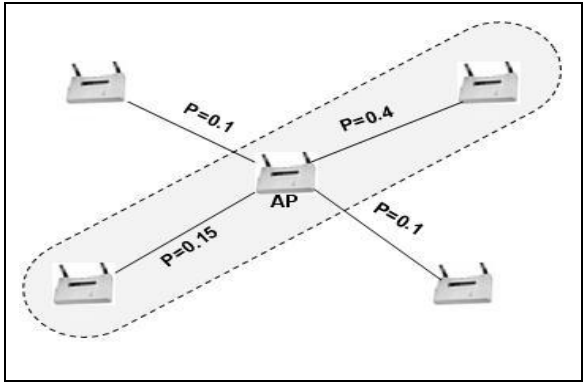

**Fig.3.** PNC and SNC scheme

These handoff probabilities are not considered in PNC. However in SNC scheme these handoff probabilities are taken in account and based on the handoff probabilities context information of a mobile node is propagated to the appropriate candidate AP. The candidate AP is selected by comparing threshold value with the handoff probabilities. If the handoff probability of an AP is equal or higher than the predefined threshold value the context will be transferred otherwise not. For example, if the threshold value is 0.15, two AP's will get the context information of the mobile node as shown in the highlighted area of Fig.3. The optimum performance of SNC scheme depends upon the predefined threshold value which should be carefully determined.

## 4   NACS (Non-overlapping AP's caching scheme)

NACS scheme works similar to the PNC scheme. AP in the PNC scheme proactively propagates the mobile node context to all neighbor AP's using the neighbor graph to capture the topological information of AP's in the wireless network. In our proposed non-overlapping APs caching scheme (NACS), mobile node context is proactively propagated to neighbor AP's which don't overlap with the current AP. It is observed, if two AP's overlap with each other than there exist a location where a MN can communicate to both of them with acceptable link quality thus the handoff can be avoided in this case. Considering this factor, we transfer the context information to those APs which do not overlap with the current AP. For the topological information of non overlapping access points non overlapping neighbor graph is generated at each AP.

A non-overlap neighbor Graph (NONG) is a complement graph of an overlap graph, meaning that $(AP_i, AP_j)$ is an edge in the non-overlap graph if and only if $(AP_i, AP_j)$ is NOT an edge in the overlap graph. The construction of NONG is much faster than NG because it does not require the mobility of station. NONG is generated by the overlapping test which can be performed with the sufficient number of mobile stations. Mobile stations do full channel scanning and reports the overlapping access points to the system. Brute-force algorithm helps us to infer the NONG.

Unlike NG, the NONG not have some missing edges but it does not have any redundant edges. Redundant edges are pair of APs that overlap each other but no handoff is possible. Non overlapping graph can be constructed by defining an undirected graph where undirected graph G = {*Vertices, Edge*} this implies *Vertices* = {$ap_a$, $ap_b$, $ap_c$ … $ap_n$} is a set of all AP's in a wireless network and there is an edge, *Edge* = ($ap_x$, $ap_y$) between $ap_x$ and $ap_y$ if they have a re-association relationship and they do not overlap with each other. Thus NON ($AP_x$) = ($ap_{xz}$) where $ap_{xz}$ ∈ *Vertices*, ($ap_x$, $ap_{xz}$) ∈ *Edge* i.e. it is a set of all non-overlapping neighbors AP's of $ap_x$ in G.

The detail mechanism of NACS is, initially non-overlapping neighbor graph (NONG) is generated at each AP. When a mobile node associates with the access point, context information of a mobile node will be propagated to all non-overlapping AP's in a wireless network. Mobile node context will be propagated to APs based on information gathered by NONG.

As a result this scheme reduces the overhead of propagating the MN context to all AP's in the network which reduces the signaling cost of context propagation. NACS diminish the overhead of calculating the handoff probabilities or neighbor weights of the AP's, which causes the low cache hit if the value of threshold is not selected.

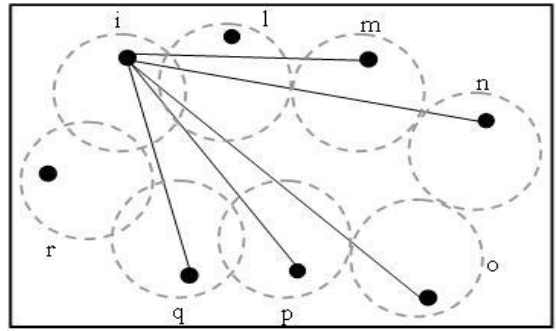

**Fig.4.** Non-Overlapping Neighboring Graph

Fig.4. shows the simulation environment. The dashed circles indicate the radio range of the APs and the black highlighted spots are referred as APs with alphabetical names for the ease of understanding. The direction of mobile node is random, that's why we described all available possibilities with respect to $AP_i$.

| **Algorithm 2** NACS ($AP_j$, c, $AP_i$) |
|---|
| Require: Algorithm executes on $ap_j$, $ap_i$ is the old AP C is the client. |
| 1.  **if** client c associates to $ap_j$ **then** |
| 2.     **for all** $ap_i$ ∈ NONG ($ap_j$) **do** |
| 3.        *Propagate Context (apj, c, api)* |
| 4.     **end for** |
| 5.  **end if** |
| 6.  **if** client c re-associate to $ap_j$  from $ap_k$ **then** |

7.  **if** Context (c) not in Cache (apj) **then**
8.  Obtain Context (ap$_j$, c, ap$_i$)
9.  **end if**
10. **for all** ap$_i$ ∈ NONG (ap$_j$) **do**
11. Propagate Context (ap$_j$, c, ap$_i$)
12. **end for**
13. **end if**
14. **if** client c re-associates to ap$_k$ from ap$_j$ **then**
15. **for all** ap$_i$ ∈ NONG (ap$_j$) **do**
16. Remove Context (ap$_j$, c, ap$_i$)
17. **end for**
18. **end if**
19. **if** ap$_j$ received Context (c) from ap$_i$ **then**
20. Insert_Cache (ap$_j$, Context(c))
21. **end if**

In NACS algorithm the context information is propagated to the non overlapping AP's (lines 2-5). If the cache context information is not found on new neighboring AP, it will request to old AP for the context information of the respective mobile node. After receiving the context information the AP$_j$ propagate the context information to all other non overlapped neighborhood AP's (lines 6-18). The AP$_i$ will remove the context information after propagation. Context information is inserted in AP cache (line 19-21).

## 5 Performance Evaluation

In proposed scheme, only non overlapping neighbor AP's receives the mobile node's context information. Therefore it will reduce the unnecessary signaling cost of propagating the mobile node context to the APs in the wireless network. Cache miss occur when a mobile node move to the AP which have not received the context information in advance. AP which have not receive the context information will perform the *obtain context ()* procedure. To evaluate the performance of our scheme, we have performed simulation for the wireless network of eight APs among them few of them have the overlapping relationship between each other as shown in Fig 4.

### 5.1 Signaling cost for Context propagation

The reduction of signaling cost is important especially when the numbers of mobile nodes increases .Signaling cost of NACS, PNC, and SNC schemes are compared in the fig 5. Signaling cost of PNC scheme is always 1 as the context is propagated to all the neighbor in the wireless network thus when the number of user increase the signaling cost vary. The signaling cost of SNC scheme is relatively less than PNC but as stated above the optimum performance of SNC is depend on the calculation of the threshold value. In NACS the context propagation cost is evaluated with the help of following equation:

Context Propagation Cost = $\sum_i \sum_j C_{ij} CP_{ij}$ (1)

Here $C_{ij}$ is the cost to transfer the context from AP$_i$ to AP$_j$ and $CP_{ij}$ is cache hit probability that the context is delivered from AP$_i$ to AP$_j$ in advance. Figure 5 shows the relative signaling cost among the NACS, SNC and PNC schemes. As a result signaling cost of the NACS scheme is significantly less then PNC and SNC scheme. Specifically NACS scheme reduces the signaling cost up to 55%

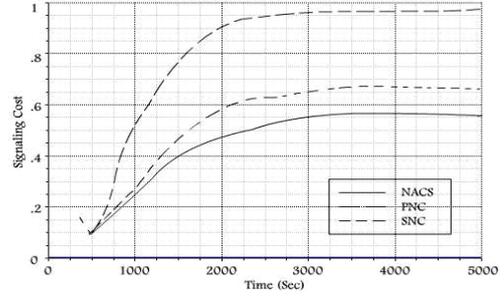

**Fig.5.** Signaling cost ratio between NACS, SNC and PNC

### 5.2 Cache Hit Probability

Increases the cache hit is parallel important issue as reducing the signaling cost of context propagation. Because if cache miss occur, the AP to which mobile node tries to associate has to communicate with the old AP to get the context information of the mobile node to re-associate it. This all process will increase the handoff latency. To calculate cache hit probability we use the following equations:

$$CHP = \frac{\sum_i \sum_j C_{hit}}{\sum_i \sum_j C_{try}} \quad (2)$$

Here $C_{hit}$ is the context searched in the cache, and $C_{try}$ is the mobile node context searched in cache.Fig.6 shows the cache hit probability of different handoff reduction schemes in different time slots. In PNC scheme, when the user mobility increase the cache hit probability also increase and it reaches almost 98%. In SNC the cache hit probability is low because the optimum performance is based on the values of threshold.

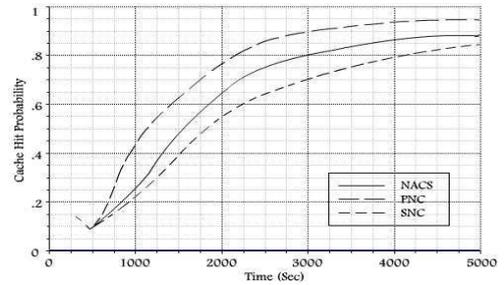

**Fig.6.** Cache Hit Probability

In NACS scheme, at the beginning, as the AP does not have the mobile node context information; the cache hit probability was low. But as the user mobility increases the cache increase like PNC scheme, the cache hit probability of the NACS scheme is almost 80%, the cache hit probability increases as the mobility of the user increases and non overlapping neighbor graph is constructed.

# 6 Conclusion

Mobility support in IEEE 802.11 is an important issue. In this paper, we have presented NACS scheme to reduce the handoff latency. We propagate mobile node context to the non overlapping access points which can be inferred by using the non overlapping neighboring graph. This graph can be generated much faster then the neighboring graph. As a result this schema reduces the signaling overhead of propagating the mobile node context to all neighbor AP. We believe that NACS is the most general and flexible context propagating schemes to date, and shows the significant potential in reducing the handoff latency.

## Reference:

[1] IEEE 802.11b WG, Part 1, "Wireless LAN medium access control (MAC) and physical layer (PHY) specification: High speed physical layer extension in 2.4 GHz Band," IEEE, September 1999.

[2] H.S. Kim, S.H. Park, C.S. Park, Jae-Won Kim, Sung-Jea Ko: Selective Channel Scanning for fast handoff in Wireless LAN using Neighbor Graph: Personal Wireless Communications: IFIP TC6 9th International Conference, PWC 2004

[3] Arunesh Mishra, Minho Shin, William Arbaugh: An Empirical Analysis of the IEEE 802.11 MAC Layer Handoff Process: ACM SIGCOMM Computer Communication Review, 2003

[4] Sangho Shin, Anshuman Singh Rawat, Henning Schulzrinne: Reducing MAC Layer Handoff Latency in IEEE 802.11 Wireless LANs: ACM SIGCOMM Computer Communication Revie, 2004

[5] Arunesh Mishra, Minho Shin, William Arbaugh: Improving the Latency of 802.11 Hand-offs using neighbor Graphs: Proceedings of the 2nd international conference on Mobile systems, applications, and services MobiSys '04

[6] Hector Velayas, Gunnar Karlsson: Techniques to Reduce the IEEE 802.11b handoff time: Apr 2003.

[7] R. Koodli and C.E. Perkins, "Fast Handover and Context Relocation in Mobile Networks," ACM SIGCOMM Computer Communication Review, vol. 31, no. 5, Oct. 2001.

[8] Arunesh Mishra, Minho Shin, William Arbaugh: Context Caching using Neighbor Graph for Fast Handoffs in a Wireless Network: INFOCOM 2004. Twenty-third AnnualJoint Conference of the IEEE Computer and Communications Societies,

[9] Sangheon Pack, Hakyung Jung, Taekyoung Kwon, and Yanghee Choi: A Selective Neighbor Caching Scheme for Fast Handoff in IEEE 802.11 Wireless Networks.

[10] JIN Xiaohui, LI Jiandong: IAPP Enhancement Protocol.

[11] Pat Calhoun and James Kempf,: Context transfer, handoff candidate discovery, and dormant mode host alerting: IETF SeaMoby Working Group.

[12] S. Pack, Y. Choi: Fast handoff scheme based on mobility prediction in public wireless LAN systems. Communications, IEE Proceedings, 2004.